\documentclass[twocolumn,showpacs,aps,pre,
groupedaddress,amssymb,amsmath,nobalancelastpage]{revtex4}

\usepackage{graphicx}
\usepackage{longtable}

\begin{document}

\title{Nonlinear stress and fluctuation dynamics of sheared disordered wet foam}

\author{Ethan Pratt}
\affiliation{Department of Physics, California Polytechnic State
University, San Luis Obispo, CA 93407}
\author{Michael Dennin}
\affiliation{Department of Physics and Astronomy and Institute for
Interfacial and Surface Science, University of California at
Irvine, Irvine, Ca 92697}

\date{\today}

\begin{abstract}

Sheared wet foam, which stores elastic energy in bubble
deformations, relaxes stress through bubble rearrangements. The
intermittency of bubble rearrangements in foam leads to
effectively stochastic drops in stress that are followed by
periods of elastic increase. We investigate global characteristics
of highly disordered foams over three decades of strain rate and
almost two decades of system size. We characterize the behavior
using a range of measures: average stress, distribution of stress
drops, rate of stress drops, and a normalized fluctuation
intensity. There is essentially no dependence on system size. As a
function of strain rate, there is a change in behavior around
shear rates of $0.07\ {\rm s^{-1}}$.

\end{abstract}


\maketitle

\section{Introduction}

One of the potentially exciting features of driven, complex fluids
is the possible existence of an ``effective'' temperature
\cite{OODLLN02, BB02}. Examples of systems for which an effective
temperature may prove a useful idea include foams, emulsion,
granular materials, and colloidal glasses (for example, see
Ref.~\cite{ITPJamming}, and references therein). Theoretical
studies of effective temperatures using the bubble model of foams
\cite{OODLLN02} and the ``standard model'' for super cooled liquid
(a binary Lennard-Jones mixture) \cite{BB02} provide strong
motivation for experimental studies of effective temperature.
Understanding the nature of fluctuations in these systems is a key
step toward developing an understanding of any concept of
effective temperature. In this paper, we focus on fluctuations in
a sheared, two-dimensional foam system: bubble rafts
\cite{AK79,LTD02}. Under shear, initially jammed foam exhibits
elastic behavior until the yield strain, followed by intermittent
``stress drops'' as the bubbles undergo nonlinear topological
rearrangements (for a review of foams see \cite{K88,S93,WH99}).
The fluctuations in stress, and other quantities in the system,
are reminiscent of thermal fluctuations and motivate definitions
of effective temperature. Previous work with bubble rafts
characterized the statistical distribution of stress drops for a
single system size and a small range of strain rates \cite{LTD02}.
The results were in excellent agreement with simulations of the
bubble model \cite{D95, D97, TSDKLL99}. In this paper, we will
report on results for a greater range of system size and strain
rate. Additionally, we will report on a number of measures other
than the distribution of stress drops. In addition to providing a
starting point for studies of effective temperatures, this work
provides detailed tests of competing models of flowing,
two-dimensional foams.

There are a number of different models of flowing foam. They all
make qualitatively similar predictions regarding the behavior of
the stress as a function of strain rate. For small
rates-of-strain, there is an initial elastic region. At a critical
value of the stress (yield stress) or strain (yield strain), the
foam begins to flow. The flow in this region is intermittent, with
periods of increase followed by sudden, irregular stress releases,
referred to as stress drops. Where the models differ is the
details of the distribution of stress drops. These differences are
the result of different assumptions concerning the source of
dissipation and dryness of the foam. As this work focuses on
two-dimensional foams, the ``dryness'' is characterized by the
area fraction of gas, $\phi$. For $\phi = 1$, foam is perfectly
dry, and the bubbles are all polygons. For $\phi < 0.84$, the foam
``melts'' into a froth of exclusively circular bubbles. Foams near
this limit are referred to as ``wet''.

Four main models of two-dimensional foam are: the vertex model
\cite{KNN89, OK95, KOKN95}; the quasi-static model
\cite{WBHA92,HWB95}; the extended q-Potts model \cite{JSSAG99};
and the bubble model \cite{D95, D97, TSDKLL99}. The vertex model
is not particularly relevant to our system, as it models dry foam.
The quasi-static model is special because it does not contain any
dissipation. It deals with wet foams. The results from this model
suggest a power-law distribution for the probability of stress
drops of a certain size occurring \cite{WBHA92,HWB95}. This result
is based mainly on measuring the distribution of T1 events. T1
events are nonlinear neighbor switching events between four
bubbles. The extended q-Potts model includes dissipation without
making any specific assumptions about the dissipation. The work in
Ref.~\cite{JSSAG99} focuses on the dry foam limit, but the model
can treat wet foams. These simulations suggest that a quasi-static
limit does not exist for foam. In other words, as the strain rate
is continually decreased, the properties of the flow continue to
change. Also, they report a transition, as a function of disorder,
from a viscoelastic solid to a viscoelastic fluid. The signature
of this change was the fact that sufficiently disordered foams
displayed no yield strain. Instead, such foams flowed immediately
upon the application of shear. Finally, this work suggests that
the distribution of energy drops in sheared foam obeys a
power-law. However, the distribution of T1 events does not. The
bubble model is applicable to wet foams, and its main prediction
with regard to stress drops is a power-law distribution for small
stress drops with an exponential cutoff at larger stress drop
magnitudes \cite{D95, D97, TSDKLL99}. In contrast to the q-Potts
model, the bubble model does predict a quasi-static limit for
flowing foams.

Experiments with foam have yielded various results. Indirect
studies of three-dimensional foam using diffusive wave
spectroscopy suggest a cutoff to the nonlinear rearrangements
\cite{GD95}. Studies of T1 events using two-dimensional foams
agree with these results \cite{DK97}. These experiments used the
gas-liquid coexistence region of Langmuir monolayers to make truly
two-dimensional foams. Experiments using quasi-static shear of a
single layer of bubbles between glass plates suggest that there
may be system-wide events, suggesting a possible power-law
behavior \cite{KE99}. Work with bubble rafts (a single layer of
bubbles on the surface of water) showed that the distribution of
stress drops for a system of approximately 900 bubbles exhibited
the exponential cutoff predicted by the bubble model \cite{LTD02}.
In order to further test the agreement with the bubble model, we
have extended the work in Ref.~\cite{LTD02} to a wide range of
system sizes and strain rates. Furthermore, in addition to
reporting on the distribution of stress drops, we also measure the
rate of stress drops, the average stress (which gives the
viscosity), and the normalized stress fluctuation. We establish
that the cutoff in stress drops is not a finite size effect. Also,
we show that for this system there is a well-defined quasi-static
limit, as predicted by the bubble model. Finally, we will discuss
the apparent dependence of some of the properties on system size
and strain rate. The rest of the paper is organized as follows.
Section II describes the experimental techniques. Section III
presents the results. Section IV discusses the results in context
of the various models of foams.

\section{Experimental Details}

\begin{figure}
\includegraphics[width=3.0in]{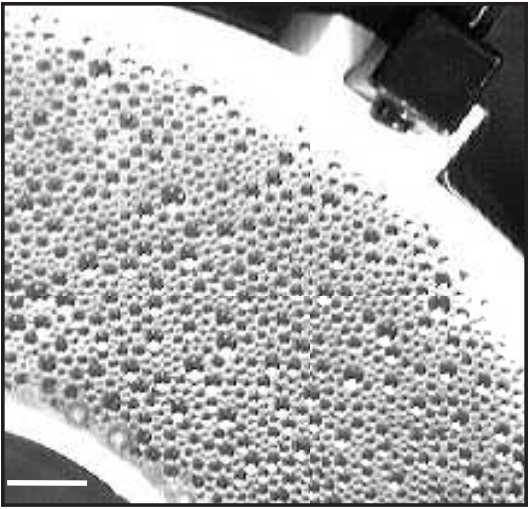}
\caption{\label{image} Image of a section of a typical bubble
raft. A portion of the inner rotor is visible in the lower left
corner. A portion of the outer barrier is visible in the upper
right corner. This particular raft had approximately $10^4$
bubbles. The scale bar is 1 cm.}
\end{figure}

A standard bubble raft \cite{AK79} provides a nice model system
for studying two-dimensional foams. Bubble rafts consist of a
layer of bubbles floating on the surface of water. The motion of
the bubbles is essentially all in the plane defined by the water
surface. However, it should be noted that the system is not an
ideal, two-dimensional system, as some motion is possible
perpendicular to the surface. However, in all of our experiments,
the bubbles were monitored with video cameras, and no motion was
observable perpendicular to the water surface. The bubble raft was
produced by flowing regulated nitrogen gas through a hypodermic
needle into a homogeneous solution of 82.0\% by volume glycerine,
14.5\% by volume deionized water, 1.50\% by volume
triethanolamine, and 2.00\% by volume oleic acid. The bubble size
was dependent on the nitrogen flow rate, which we varied using a
needle valve. The bubble diameter ranged from 2 to 6 mm, with most
bubbles in the 3 to 4 mm range. The resulting bubbles were spooned
into a cylindrical Couette viscometer described in detail in
Ref.~\cite{app}. This produced a two-dimensional, wet foam on a
homogeneous liquid substrate of 80\% by volume deionized water,
15\% by volume glycerine, and 5.0\% by volume Miracle Bubbles
(Imperial Toy Corp.) The Couette viscometer consists of a shallow
dish that contains the liquid substrate. Two concentric Teflon
rings are placed vertically in the dish. The outer ring consists
of twelve segmented pieces and has an adjustable radius. The inner
ring, or rotor, has a radius $r = 4.0\ {\rm cm}$ and was suspended
by a wire to form a torsion pendulum. Polypropylene balls with a 4
mm diameter were epoxied to the circumference of the inner rotor
to prevent the innermost row of bubbles from slipping. The
outermost row of bubbles was monitored, and no slip of bubbles on
the outer barrier was observed. Figure~\ref{image} is a top view
of a typical bubble raft in our apparatus. Portions of the two
Teflon rings are visible.

To shear the foam, the outer Teflon barrier was rotated at a
constant angular velocity in the range 0.0005 rad/s to 0.5 rad/s.
The torque, $\tau$, on the inner rotor was monitored by recording
the angular position of the inner rotor (and thus the angular
displacement of the torsion wire) twice per second. The angular
position was measured using magnetic flux techniques described in
detail in Ref.~\cite{app}. This data was recorded by a 12-bit
analog-to-digital converter and stored on a PC. The tangential
stress, $\sigma$, on the inner rotor due to the foam is given by
$\sigma = \tau/(2\pi r^2)$. During periods without any
rearrangements, the fluctuations in the stress were at the level
of one bit, corresponding to changes in stress of $2 \times
10^{-3}\ {\rm dyne/cm}$. Therefore, when computing stress drops,
we filtered the data to eliminate any changes in stress of $\pm 2
\times 10^{-3}\ {\rm dyne/cm}$.

The bubble raft was constructed by placing the approximate number
of desired bubbles in the trough with the outer barrier set to a
large radius. It is important to note that the bubbles exhibited a
strong attraction to each other. This is a microscopic detail that
is not included in any of the models discussed in the
introduction. Typically, two-dimensional foams are characterized
by their gas-area fraction, which is the ratio of the area filled
by gas to the total area. Because the bubbles actually exist in
three dimensions for a bubble raft, the fluid walls (and the
cross-sectional area of a bubble) are height dependent. This
complicates the definition of gas-area fraction. Therefore, we
used a functional definition for gas-area fraction based on the
images of the bubble raft. We defined the area of gas to be the
black regions within bubbles in an image and maintained constant
lighting conditions so that this definition was consistent from
run to run. The outer barrier was compressed until the desired
bubble density was achieved. It should be noted that this resulted
in a variation in the initial shear stress of the bubble raft that
did not relax significantly on the time-scale of the experiments.
Therefore, due to the finite lifetime of the raft, the experiments
were carried out with this initial pre-stress present. Both the
total number of bubbles and average gas area fraction were
determined from images of a large section of the trough, assuming
an essentially uniform distribution of bubbles throughout the
trough. For all of the data reported here, the gas area fraction
was approximately 0.95.

The stability of the bubbles was enhanced by cooling the fluid
substrate to $5\ ^{\circ}{\rm C}$. Also, a glass cover was placed
over the bubbles. The cover helped reduce evaporation and was not
in contact with the bubbles. The entire apparatus was contained in
a cabinet. The cabinet reduced the air flow around the apparatus,
and a humidifier placed within the cabinet helped to extend the
lifetime of the bubbles. The bubbles in the bubble raft did not
exhibit any substantial coarsening with time. Instead, the raft
tended to suffer catastrophic failure after approximately two
hours due to a significant number of bubbles popping. Presumably,
this was due primarily to the loss of fluid in the bubble walls
from drainage into the fluid substrate and/or evaporation. Because
the bubble raft did not coarsen significantly, there is no
competition between coarsening and shear induced rearrangements.

\section {Experimental Results}

\begin{figure}
\includegraphics[width=3.0in]{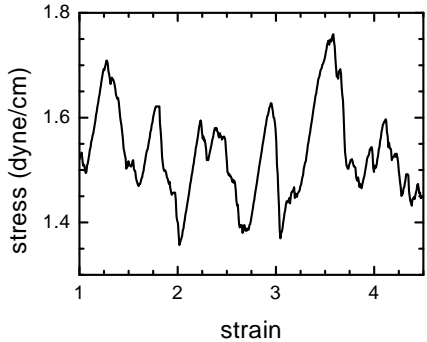}
\caption{\label{stressdrop} Typical stress response of a sheared
wet foam illustrating the intermittent stress drops. The rate of
strain is $0.014\ {\rm s^{-1}}$, and the number of bubbles is $1.6
\times 10^3$.}
\end{figure}

\begin{figure}
\includegraphics[width=3.0in]{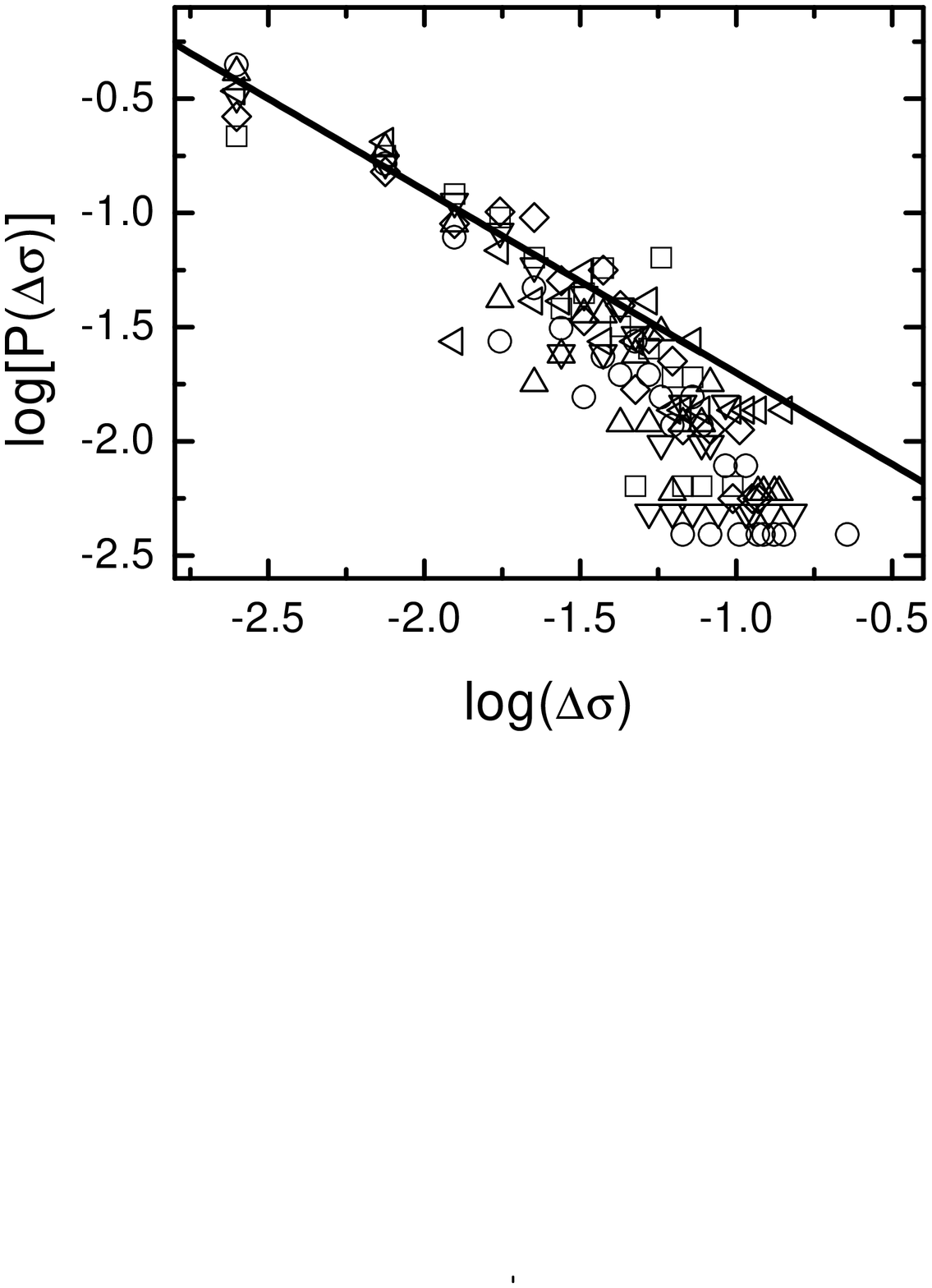}
\caption{\label{dropdist} Probability distribution for stress
drops ($P(\Delta\sigma)$) as a function of the magnitude of the
stress drop ($\Delta\sigma \equiv \delta\sigma/\sigma_{max}$). For
each separate run, the stress drops ($\delta\sigma$) have been
normalized by the maximum stress ($\sigma_{max}$). The symbols are
different strain rates: $2.7 \times 10^{-3}\ {\rm s^{-1}}$
($\bigtriangledown$); $1.4 \times 10^{-3}\ {\rm s^{-1}}$
($\bigtriangleup$); $7 \times 10^{-3}\ {\rm s^{-1}}$ ($\lhd$);
$1.4 \times 10^{-2}\ {\rm s^{-1}}$ ($\circ$); $1.3 \times 10^{-1}\
{\rm s^{-1}}$ ($\Diamond$); $2.7 \times 10^{-1}\ {\rm s^{-1}}$
($\Box$). The solid line is a guide to the eye and has a slope of
-0.8}
\end{figure}

Figure~\ref{stressdrop} shows the typical behavior of the stress
versus strain for strains above the yield strain. The irregular
behavior of the stress during flow is apparent. This behavior can
be characterized by considering the distribution of stress drops.
This is shown in Fig.~\ref{dropdist} for a system with 1550
bubbles for various strain rates ($\dot{\gamma}$). All results are
given in terms of normalized stress drops ($\Delta\sigma \equiv
\delta\sigma/\sigma_{max}$), where the stress drops $\delta\sigma$
for each run are normalized by the maximum stress ($\sigma_{max}$)
for that run. A bin size for $\Delta\sigma$ of $2.5 \times
10^{-3}$ was used in plotting the distribution, with the
probability of a stress drop of size $\Delta\sigma$
($P(\Delta\sigma)$) defined as the number of drops within each bin
divided by the total number of drops in the run. The solid line is
a guide to the eye and has a slope of -0.8. The distribution is
consistent with a power law for small stress drops with an
exponential cutoff. Because of the cutoff, there is a a
well-defined average stress drop, $<\Delta\sigma>$. The
$P(\Delta\sigma)$ for different system sizes are qualitatively the
same, a power-law for small $\Delta\sigma$, with an exponential
cutoff. In order to look for quantitative differences as a
function of system size, we considered the behavior of
$<\Delta\sigma>$.

The average stress drop is shown as a function of strain rate for
different system sizes in Fig.~\ref{avesize}. The smallest system
consisted of $1.6 \times 10^3$ bubbles and is given by the
squares. For this system, there is essentially no dependence of
$<\Delta\sigma>$ on strain rate. For the systems with more than
$5.6 \times 10^3$ bubbles (all of the other systems that we
studied), there does appear to be a weak strain rate dependence.
One observes an increase in the average stress drop with strain
rate, until $\dot{\gamma} \approx 0.07\ {\rm s^{-1}}$. Above this
value, the average stress-drop is independent of strain rate. The
implications of this will be discussion in Sec. IV.

One important result is that there is no increase in
$<\Delta\sigma>$ with system size. This effectively rules out
system size as the source of the cutoff. At the lowest strain
rates, $<\Delta\sigma>$ is lower for the larger systems. A
possible reason for this behavior is discussed in Sec. IV.

\begin{figure}
\includegraphics[width=3.0in]{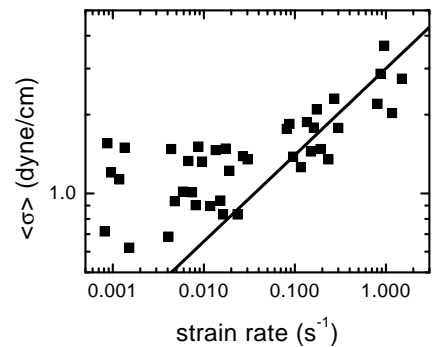}
\caption{\label{stress-strain} Plot of the average stress versus
the strain rate. The dark line is a guide to the eye, and has
slope 1/3.}
\end{figure}

The relation between average stress and strain rate is shown in
Fig.~\ref{stress-strain}. The maximum stress displays a similar
dependence. Note the knee of the curve, at $\dot{\gamma} \approx
0.07\ {\rm s^{-1}}$. A line is drawn as a guide to the eye, with a
slope of 1/3. The overall curve is consistent with a
Herschel-Bulkley model of viscosity where the stress is given by
$\sigma = A + B \dot{\gamma}^n$ \cite{BAH77}. Here $\dot{\gamma}$
is the strain rate and A and B are constants. The variation in
average stress in Fig.~\ref{stress-strain} is most likely due to
the variation in the initial stress from run to run discussed in
Sec. II. Presumably, if sufficient aging of the system were
possible before each run, this variation would be reduced.

\begin{figure}
\includegraphics[width=3.0in]{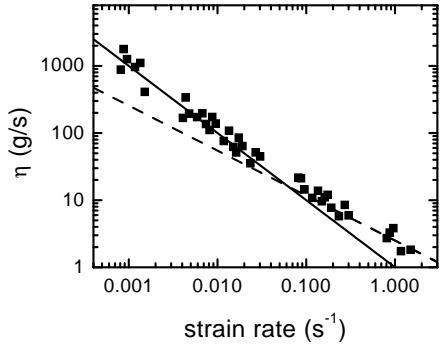}
\caption{\label{viscosity} Plot of average stress divided by the
rate of strain ($\eta$) versus the rate of strain. The solid line
has a slope of -1 and the dashed line has a slope of -2/3. The two
lines cross at $\dot{\gamma} = 0.066\ {\rm s^{-1}}$.}
\end{figure}

An alternate way to view the same data is to consider $\eta \equiv
\ <\sigma>/\dot{\gamma}$ versus $\dot{\gamma}$, shown in
Fig.~\ref{viscosity}. This is the steady-state viscosity, taking
care in computing $\dot{\gamma}$ at the inner cylinder
\cite{BAH77}. The solid line with slope -1 and dashed line with
slope -2/3 are guides to the eye and clearly illustrate the
Herschel-Bulkley behavior of the bubble raft, with an exponent $n
= 1/3$. This behavior is consistent with the shear-thinning
velocity profile reported in Ref.~\cite{LTD02} for $\dot{\gamma} =
0.062\ {\rm s^{-1}}$. An open question is the behavior of the
velocity profile at extremely low strain rates, where the average
stress is essentially independent of strain rate. One definitely
observes bubble motions throughout the bulk of the system.
However, at these low shear rates, most of the time the system is
undergoing a linear increase of the stress, and only occasionally
is there a stress drop. Initial measurements of the flow during
such an increase in the stress are consistent with a linear
profile of the velocity. However, detailed measurements in this
regime will be conducted in the future to determine what the
long-time average (one that includes many stress drops) of the
flow profile is. This is important given measurements of velocity
profiles at low shear rates that report an exponential decay of
the velocity for a similar system \cite{DTM01}. For our system,
one issue is whether or not the water substrate dragged the
bubbles. We made a number of measurements where the outer cylinder
was rotated, but the bubble raft was not in contact with the outer
cylinder. This was accomplished by removing the approximately the
outer three rows of bubbles. Under these conditions, no flow of
the bubble raft was observed, and no measurable stress was
transmitted to the inner rotor. This provides strong evidence that
the underlying water does not ``drag'' the bubble raft.

\begin{figure}
\includegraphics[width=3.0in]{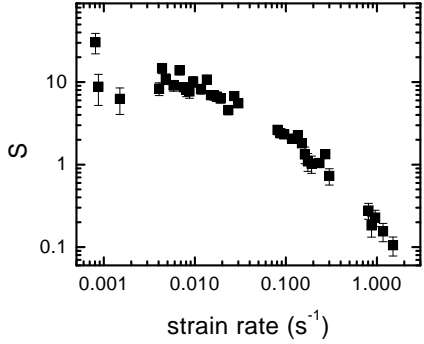}
\caption{\label{droprate} The number of stress drops per unit
strain ($S$) as a function of strain rate. The plateau at low
rates of strain suggests a quasi-static limit is reached.}
\end{figure}

The dependence of the maximum stress on strain rate suggests the
existence of a quasi-static limit. Below a strain rate of
approximately $0.07\ {\rm s^{-1}}$, the stress is essentially
independent of the rate of strain. This was checked by considering
the number of stress drops per unit strain ($S$). This is plotted
in Fig.~\ref{droprate}. As with the maximum stress, $S$ approaches
a constant below values of the strain rate of approximately $0.07\
{\rm s^{-1}}$.

\begin{figure}
\includegraphics[width=3.0in]{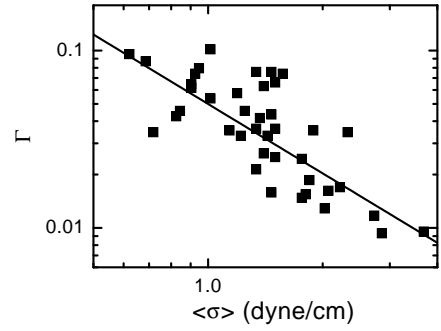}
\caption{\label{flucstress} The fluctuation intensity ($\Gamma$)
versus the mean stress. The dark line is a guide to the eye, and
has slope -1.3.}
\end{figure}

In addition to the distribution of stress drops, we also
characterized the intensity of fluctuations around the mean. We
defined the fluctuation intensity, $\Gamma$, as the standard
deviation of the stress (after the yield stress) for a given run,
expressed as a fraction of the mean stress:
\begin{equation}
\Gamma \equiv \sqrt{\frac{1}{N}\sum_{i=1}^{N}\left(\frac{\sigma_i
- <\sigma>}{<\sigma>}\right)^2}
\end{equation}
where the sum is over the measured values of stress and N is the
number of data points for a given experimental run. Given the
existence of the pre-stress, the normalization by the average
stress allows for a better comparison between different systems.
Figure~\ref{fluctinten} shows the results for the fluctuation
intensity ($\Gamma$) as a function of $\dot{\gamma}$ for the
different system sizes. There is significant scatter in the data;
however, there is a clear trend of decreasing $\Gamma$ as
$\dot{\gamma}$ increases. One consequence of this is a correlation
between $\Gamma$ and the $<\sigma>$. This is illustrated in
Fig.~\ref{flucstress}, where the data for the different system
sizes are combined into a single plot. Again, there is significant
scatter in the data, but the trend is obvious.

\section{Summary}

The results presented here provide strong evidence that the bubble
model provides an accurate description of the shear behavior of a
bubble raft. To the extent that bubble rafts are equivalent to
foam, the bubble model would also describe a two-dimensional
flowing foam. The stress drop distribution, the average stress as
a function of strain rate (the steady state viscosity), and the
rate of stress drops, $S$, are all consistent with the bubble
model \cite{TSDKLL99}. This agreement is despite the fact that the
bubble rafts studied here are strongly attractive, a feature that
is not explicitly in the model. One interesting result is that
both the bubble model and the bubble rafts are well-described as a
Herschel-Bulkley fluid with an exponent of 1/3 \cite{LL00}. It
remains to be seen if this exponent is a generic feature of the
model, and of similar models, or if there is something specific to
the parameters used in Ref.~\cite{LL00}. For example, the exponent
may depend on various characteristics of the foam, such as the
gas-area fraction. This dependence also needs to be tested for the
bubble rafts.

Comparison of our studies with both the quasi-static model
\cite{WBHA92,HWB95} and the quasi-static experiments \cite{KE99}
raises an interesting question: is there a fundamental difference
between slow but steady shear, and true quasi-static motions? The
disagreement between our results and the quasi-static experiments
suggests that such a difference may exist. However, it is also
possible that the discrepancies are due to comparing direct
measurements of the stress drops with sizes of spatial
rearrangements. Future work with our system will look at both the
issue of quasi-static steps versus steady shear and the spatial
extent of rearrangements.

With regard to the extended q-Potts model, this work raises some
important questions. Two clear predictions of that model are: (1)
there is no quasi-static limit; and (2) a sufficiently disordered
foam no longer has a yield strain \cite{JSSAG99}. Neither behavior
was observed in our experiments. At this point, one would need to
do further work to determine if there was something fundamentally
missing from the q-Potts model that results in this disagreement.
The other possibility is that our foams were either not
sufficiently disordered to be accurately described by the q-Potts
model or they were not sufficiently dry, as the simulations in
Ref.~\cite{JSSAG99} were for dry foam. Therefore, future
experiments will focus on the role of disorder and the wetness of
the foam.

\begin{figure}
\includegraphics[width=3.0in]{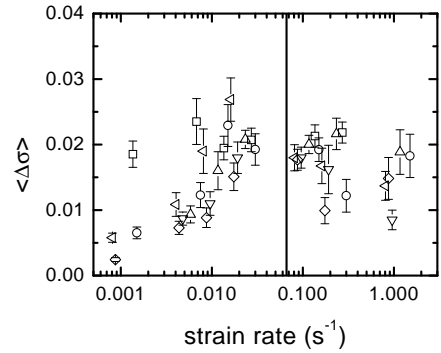}
\caption{\label{avesize} Plot of the average size of the stress
drop ($<\Delta\sigma>$) as a function of the strain rate for
different system sizes: $1.6 \times 10^3$ ($\Box$); $5.6 \times
10^3$ ($\circ$); $9.2 \times 10^3$ ($\bigtriangleup$); $1.5 \times
10^4$ ($\bigtriangledown$); $2.0 \times 10^4$ ($\Diamond$); $2.6
\times 10^4$ ($\lhd$). The solid vertical line is at $\dot{\gamma}
= 0.066\ {\rm s^{-1}}$.}
\end{figure}

Though not conclusive, the behavior of the average stress drop as
a function of system size and strain rate, shown in
Fig.~\ref{avesize}, suggests some interesting behavior. The large
decrease in $<\Delta\sigma>$ as a function of system size is
surprising. One possible explanation involves the spatial
correlations between bubble rearrangements that produce the stress
drops. As the system size increases, there are more spatial
locations at which rearrangements can occur. For low enough strain
rates, there will be an intermediate range of systems size for
which this increases the probability of isolated small stress
drops occurring. Once one region slips, enough stress is relieved
that the other regions do not rearrange until a sufficiently later
time that they are recorded as a new stress drop. Such dynamics
would result in a decrease in the average stress drop with system
size. Eventually, as the system size increases even more, this
behavior should ``smooth'' out the dynamics, as small stress drops
occur almost continuously. Presumably, this happens in large,
three-dimensional samples. On the other hand, for these
intermediate size systems, as the strain rate is increased, the
stress releases occur closer together. This increases the
likelihood of multiple small events in different spatial locations
combining to form larger stress drops. Therefore, one observes an
increase in the average stress drop as a function of strain rate.
Eventually, as the system crosses over to more fluid-like
behavior, there is again a ``smoothing'' of the dynamics. In this
regime, the average stress drop becomes independent of strain
rate. Clearly, more work is needed, both in the experiments and
simulations, to test these ideas. In particular, they highlight
the importance of measuring both spatial correlations between
rearrangement events and the correlation between the
rearrangements and the stress drops.

\begin{figure}
\includegraphics[width=3.0in]{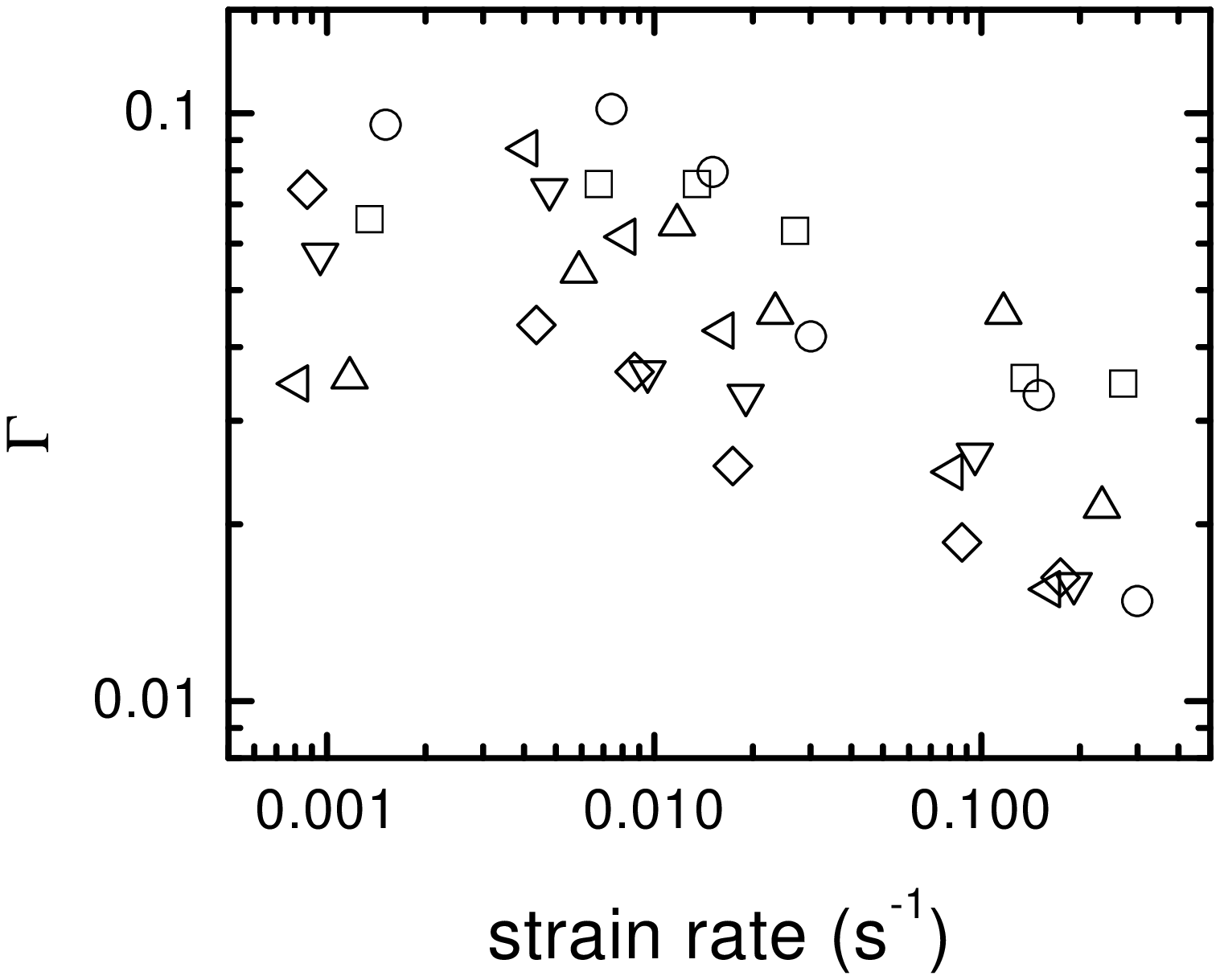}
\caption{\label{fluctinten}. Fluctuation intensity ($\Gamma$)
versus strain rate for a range of system sizes: $1.6 \times 10^3$
($\Box$); $5.6 \times 10^3$ ($\circ$); $9.2 \times 10^3$
($\bigtriangleup$); $1.5 \times 10^4$ ($\bigtriangledown$); $2.0
\times 10^4$ ($\Diamond$); $2.6 \times 10^4$ ($\lhd$).}
\end{figure}

Finally, it interesting that the crossover to ``smoother,'' more
fluid-like behavior, as a function of strain rate, is evident in
both the measurement of $\Gamma$ and $<\Delta\sigma>$. However,
the two measures reveal slightly different behavior. As one
increases the strain rate, $\Gamma$ decreases monotonically (see
Fig.~\ref{fluctinten}). However, $<\Delta\sigma>$ becomes
independent of $\dot{\gamma}$ at the higher strain rates (see
Fig.~\ref{avesize}). The two results are not inconsistent, as
$\Gamma$ measures the size of fluctuations from the mean, while
$<\Delta\sigma>$ measures the average size of the changes in
stress that produce these fluctuations. Again, understanding the
spatial distribution of bubble rearrangements will probably be an
important step in fully understanding the dependence of these two
measures of the fluctuations on strain rate.

\begin{acknowledgements}

The authors acknowledge funding from NSF grant CTS-0085751. M.
Dennin acknowledges further funding from the Research Corporation
and Alfred P. Sloan Foundation. E. Pratt was supported by NSF
Research Experience for Undergraduates (REU) grant PHY-9988066.
The authors thank A. J. Liu and C. O'Hern for useful discussion of
the bubble model and Y. Jiang for discussions of the q-Potts
model.

\end{acknowledgements}


\begin{thebibliography}{24}
\expandafter\ifx\csname
natexlab\endcsname\relax\def\natexlab#1{#1}\fi
\expandafter\ifx\csname bibnamefont\endcsname\relax
  \def\bibnamefont#1{#1}\fi
\expandafter\ifx\csname bibfnamefont\endcsname\relax
  \def\bibfnamefont#1{#1}\fi
\expandafter\ifx\csname citenamefont\endcsname\relax
  \def\citenamefont#1{#1}\fi
\expandafter\ifx\csname url\endcsname\relax
  \def\url#1{\texttt{#1}}\fi
\expandafter\ifx\csname
urlprefix\endcsname\relax\def\urlprefix{URL }\fi
\providecommand{\bibinfo}[2]{#2}
\providecommand{\eprint}[2][]{\url{#2}}

\bibitem[{\citenamefont{Ono et~al.}(2002)\citenamefont{Ono, O'Hern, Durian,
  Langer, Liu, and Nagel}}]{OODLLN02}
\bibinfo{author}{\bibfnamefont{I.~K.} \bibnamefont{Ono}},
  \bibinfo{author}{\bibfnamefont{C.~S.} \bibnamefont{O'Hern}},
  \bibinfo{author}{\bibfnamefont{D.~J.} \bibnamefont{Durian}},
  \bibinfo{author}{\bibfnamefont{S.~A.} \bibnamefont{Langer}},
  \bibinfo{author}{\bibfnamefont{A.~J.} \bibnamefont{Liu}}, \bibnamefont{and}
  \bibinfo{author}{\bibfnamefont{S.~R.} \bibnamefont{Nagel}},
  \bibinfo{journal}{Phys. Rev. Lett.} \textbf{\bibinfo{volume}{89}},
  \bibinfo{pages}{095703} (\bibinfo{year}{2002}).

\bibitem[{\citenamefont{Berthier and Barrat}(2002)}]{BB02}
\bibinfo{author}{\bibfnamefont{L.}~\bibnamefont{Berthier}} \bibnamefont{and}
  \bibinfo{author}{\bibfnamefont{J.-L.} \bibnamefont{Barrat}},
  \bibinfo{journal}{Phys. Rev. Lett.} \textbf{\bibinfo{volume}{89}},
  \bibinfo{pages}{095702} (\bibinfo{year}{2002}).

\bibitem[{\citenamefont{Liu and Nagel}(2001)}]{ITPJamming}
\bibinfo{editor}{\bibfnamefont{A.~J.} \bibnamefont{Liu}} \bibnamefont{and}
  \bibinfo{editor}{\bibfnamefont{S.~R.} \bibnamefont{Nagel}}, eds.,
  \emph{\bibinfo{title}{Jamming and Rheology}} (\bibinfo{publisher}{Taylor and
  Francis Group}, \bibinfo{year}{2001}).

\bibitem[{\citenamefont{Argon and Kuo}(1979)}]{AK79}
\bibinfo{author}{\bibfnamefont{A.~S.} \bibnamefont{Argon}} \bibnamefont{and}
  \bibinfo{author}{\bibfnamefont{H.~Y.} \bibnamefont{Kuo}},
  \bibinfo{journal}{Mat. Sci. and Eng.} \textbf{\bibinfo{volume}{39}},
  \bibinfo{pages}{101} (\bibinfo{year}{1979}).

\bibitem[{\citenamefont{Lauridsen et~al.}(2002)\citenamefont{Lauridsen,
  Twardos, and Dennin}}]{LTD02}
\bibinfo{author}{\bibfnamefont{J.}~\bibnamefont{Lauridsen}},
  \bibinfo{author}{\bibfnamefont{M.}~\bibnamefont{Twardos}}, \bibnamefont{and}
  \bibinfo{author}{\bibfnamefont{M.}~\bibnamefont{Dennin}},
  \bibinfo{journal}{Phys. Rev. Lett.} \textbf{\bibinfo{volume}{89}},
  \bibinfo{pages}{098303} (\bibinfo{year}{2002}).

\bibitem[{\citenamefont{Kraynik}(1988)}]{K88}
\bibinfo{author}{\bibfnamefont{A.~M.} \bibnamefont{Kraynik}},
  \bibinfo{journal}{Ann. Rev. Fluid Mech.} \textbf{\bibinfo{volume}{20}},
  \bibinfo{pages}{325} (\bibinfo{year}{1988}).

\bibitem[{\citenamefont{Stavans}(1993)}]{S93}
\bibinfo{author}{\bibfnamefont{J.}~\bibnamefont{Stavans}},
  \bibinfo{journal}{Rep. Prog. Phys.} \textbf{\bibinfo{volume}{56}},
  \bibinfo{pages}{733} (\bibinfo{year}{1993}).

\bibitem[{\citenamefont{Weaire and Hutzler}(1999)}]{WH99}
\bibinfo{author}{\bibfnamefont{D.}~\bibnamefont{Weaire}} \bibnamefont{and}
  \bibinfo{author}{\bibfnamefont{S.}~\bibnamefont{Hutzler}},
  \emph{\bibinfo{title}{The Physics of Foams}} (\bibinfo{publisher}{Claredon
  Press, Oxford}, \bibinfo{year}{1999}).

\bibitem[{\citenamefont{Durian}(1995)}]{D95}
\bibinfo{author}{\bibfnamefont{D.~J.} \bibnamefont{Durian}},
  \bibinfo{journal}{Phys. Rev. Lett.} \textbf{\bibinfo{volume}{75}},
  \bibinfo{pages}{4780} (\bibinfo{year}{1995}).

\bibitem[{\citenamefont{Durian}(1997)}]{D97}
\bibinfo{author}{\bibfnamefont{D.~J.} \bibnamefont{Durian}},
  \bibinfo{journal}{Phys. Rev. E} \textbf{\bibinfo{volume}{55}},
  \bibinfo{pages}{1739} (\bibinfo{year}{1997}).

\bibitem[{\citenamefont{Tewari et~al.}(1999)\citenamefont{Tewari, Schiemann,
  Durian, Knobler, Langer, and Liu}}]{TSDKLL99}
\bibinfo{author}{\bibfnamefont{S.}~\bibnamefont{Tewari}},
  \bibinfo{author}{\bibfnamefont{D.}~\bibnamefont{Schiemann}},
  \bibinfo{author}{\bibfnamefont{D.~J.} \bibnamefont{Durian}},
  \bibinfo{author}{\bibfnamefont{C.~M.} \bibnamefont{Knobler}},
  \bibinfo{author}{\bibfnamefont{S.~A.} \bibnamefont{Langer}},
  \bibnamefont{and} \bibinfo{author}{\bibfnamefont{A.~J.} \bibnamefont{Liu}},
  \bibinfo{journal}{Phys. Rev. E} \textbf{\bibinfo{volume}{60}},
  \bibinfo{pages}{4385} (\bibinfo{year}{1999}).

\bibitem[{\citenamefont{Kawasaki et~al.}(1989)\citenamefont{Kawasaki, Nagai,
  and Nakashima}}]{KNN89}
\bibinfo{author}{\bibfnamefont{K.}~\bibnamefont{Kawasaki}},
  \bibinfo{author}{\bibfnamefont{T.}~\bibnamefont{Nagai}}, \bibnamefont{and}
  \bibinfo{author}{\bibfnamefont{K.}~\bibnamefont{Nakashima}},
  \bibinfo{journal}{Phil. Mag. B} \textbf{\bibinfo{volume}{60}},
  \bibinfo{pages}{399} (\bibinfo{year}{1989}).

\bibitem[{\citenamefont{Okuzono and Kawasaki}(1995)}]{OK95}
\bibinfo{author}{\bibfnamefont{T.}~\bibnamefont{Okuzono}} \bibnamefont{and}
  \bibinfo{author}{\bibfnamefont{K.}~\bibnamefont{Kawasaki}},
  \bibinfo{journal}{Phys. Rev. E} \textbf{\bibinfo{volume}{51}},
  \bibinfo{pages}{1246} (\bibinfo{year}{1995}).

\bibitem[{\citenamefont{Kawasaki et~al.}(1992)\citenamefont{Kawasaki, Okuzono,
  Kawakatsu, and Nagai}}]{KOKN95}
\bibinfo{author}{\bibfnamefont{K.}~\bibnamefont{Kawasaki}},
  \bibinfo{author}{\bibfnamefont{T.}~\bibnamefont{Okuzono}},
  \bibinfo{author}{\bibfnamefont{T.}~\bibnamefont{Kawakatsu}},
  \bibnamefont{and} \bibinfo{author}{\bibfnamefont{T.}~\bibnamefont{Nagai}}, in
  \emph{\bibinfo{booktitle}{Proc. Int. Workshop of Physics of Pattern
  Formation}}, edited by \bibinfo{editor}{\bibfnamefont{S.}~\bibnamefont{Kai}}
  (\bibinfo{publisher}{Singapore: World Scientific}, \bibinfo{year}{1992}).

\bibitem[{\citenamefont{Weaire et~al.}(1992)\citenamefont{Weaire, Bolton,
  Herdtle, and Aref}}]{WBHA92}
\bibinfo{author}{\bibfnamefont{D.}~\bibnamefont{Weaire}},
  \bibinfo{author}{\bibfnamefont{F.}~\bibnamefont{Bolton}},
  \bibinfo{author}{\bibfnamefont{T.}~\bibnamefont{Herdtle}}, \bibnamefont{and}
  \bibinfo{author}{\bibfnamefont{H.}~\bibnamefont{Aref}},
  \bibinfo{journal}{Phil. Mag. Lett.} \textbf{\bibinfo{volume}{66}},
  \bibinfo{pages}{293} (\bibinfo{year}{1992}).

\bibitem[{\citenamefont{Hutzler et~al.}(1995)\citenamefont{Hutzler, Weaire, and
  Bolton}}]{HWB95}
\bibinfo{author}{\bibfnamefont{S.}~\bibnamefont{Hutzler}},
  \bibinfo{author}{\bibfnamefont{D.}~\bibnamefont{Weaire}}, \bibnamefont{and}
  \bibinfo{author}{\bibfnamefont{F.}~\bibnamefont{Bolton}},
  \bibinfo{journal}{Phil. Mag. B} \textbf{\bibinfo{volume}{71}},
  \bibinfo{pages}{277} (\bibinfo{year}{1995}).

\bibitem[{\citenamefont{Jiang et~al.}(1999)\citenamefont{Jiang, Swart, Saxena,
  Asipauskas, and Glazier}}]{JSSAG99}
\bibinfo{author}{\bibfnamefont{Y.}~\bibnamefont{Jiang}},
  \bibinfo{author}{\bibfnamefont{P.~J.} \bibnamefont{Swart}},
  \bibinfo{author}{\bibfnamefont{A.}~\bibnamefont{Saxena}},
  \bibinfo{author}{\bibfnamefont{M.}~\bibnamefont{Asipauskas}},
  \bibnamefont{and} \bibinfo{author}{\bibfnamefont{J.~A.}
  \bibnamefont{Glazier}}, \bibinfo{journal}{Phys. Rev. E}
  \textbf{\bibinfo{volume}{59}}, \bibinfo{pages}{5819} (\bibinfo{year}{1999}).

\bibitem[{\citenamefont{Gopal and Durian}(1995)}]{GD95}
\bibinfo{author}{\bibfnamefont{A.~D.} \bibnamefont{Gopal}} \bibnamefont{and}
  \bibinfo{author}{\bibfnamefont{D.~J.} \bibnamefont{Durian}},
  \bibinfo{journal}{Phys. Rev. Lett.} \textbf{\bibinfo{volume}{75}},
  \bibinfo{pages}{2610} (\bibinfo{year}{1995}).

\bibitem[{\citenamefont{Dennin and Knobler}(1997)}]{DK97}
\bibinfo{author}{\bibfnamefont{M.}~\bibnamefont{Dennin}} \bibnamefont{and}
  \bibinfo{author}{\bibfnamefont{C.~M.} \bibnamefont{Knobler}},
  \bibinfo{journal}{Phys. Rev. Lett.} \textbf{\bibinfo{volume}{78}},
  \bibinfo{pages}{2485} (\bibinfo{year}{1997}).

\bibitem[{\citenamefont{Kader and Earnshaw}(1999)}]{KE99}
\bibinfo{author}{\bibfnamefont{A.~A.} \bibnamefont{Kader}} \bibnamefont{and}
  \bibinfo{author}{\bibfnamefont{J.~C.} \bibnamefont{Earnshaw}},
  \bibinfo{journal}{Phys. Rev. Lett.} \textbf{\bibinfo{volume}{82}},
  \bibinfo{pages}{2610} (\bibinfo{year}{1999}).

\bibitem[{\citenamefont{Ghaskadvi and Dennin}(1998)}]{app}
\bibinfo{author}{\bibfnamefont{R.~S.} \bibnamefont{Ghaskadvi}}
  \bibnamefont{and} \bibinfo{author}{\bibfnamefont{M.}~\bibnamefont{Dennin}},
  \bibinfo{journal}{Rev. Sci. Instr.} \textbf{\bibinfo{volume}{69}},
  \bibinfo{pages}{3568} (\bibinfo{year}{1998}).

\bibitem[{\citenamefont{Bird et~al.}(1977)\citenamefont{Bird, Armstrong, and
  Hassuage}}]{BAH77}
\bibinfo{author}{\bibfnamefont{R.~B.} \bibnamefont{Bird}},
  \bibinfo{author}{\bibfnamefont{R.~C.} \bibnamefont{Armstrong}},
  \bibnamefont{and} \bibinfo{author}{\bibfnamefont{O.}~\bibnamefont{Hassuage}},
  \emph{\bibinfo{title}{Dynamics of Polymer Liquids}}
  (\bibinfo{publisher}{Wiley, New York}, \bibinfo{year}{1977}).

\bibitem[{\citenamefont{Debr\'{e}geas et~al.}(2001)\citenamefont{Debr\'{e}geas,
  Tabuteau, and di~Meglio}}]{DTM01}
\bibinfo{author}{\bibfnamefont{G.}~\bibnamefont{Debr\'{e}geas}},
  \bibinfo{author}{\bibfnamefont{H.}~\bibnamefont{Tabuteau}}, \bibnamefont{and}
  \bibinfo{author}{\bibfnamefont{J.~M.} \bibnamefont{di~Meglio}},
  \bibinfo{journal}{Phys. Rev. Lett.} \textbf{\bibinfo{volume}{87}},
  \bibinfo{pages}{178305} (\bibinfo{year}{2001}).

\bibitem[{\citenamefont{Langer and Liu}(2000)}]{LL00}
\bibinfo{author}{\bibfnamefont{S.~A.} \bibnamefont{Langer}} \bibnamefont{and}
  \bibinfo{author}{\bibfnamefont{A.~J.} \bibnamefont{Liu}},
  \bibinfo{journal}{Europhys. Lett.} \textbf{\bibinfo{volume}{49}},
  \bibinfo{pages}{68} (\bibinfo{year}{2000}).

\end{thebibliography}

\clearpage

\end{document}